\documentclass[conference]{IEEEtran}
\IEEEoverridecommandlockouts

\usepackage{amsmath,amsfonts}
\usepackage{mathtools, bm}
\usepackage{algorithmic}
\usepackage[linesnumbered,ruled,vlined]{algorithm2e} % core for \KwIn, \tcp, \If, etc.
\usepackage{array}
\usepackage[caption=false,font=normalsize,labelfont=sf,textfont=sf]{subfig}
\usepackage{textcomp}
\usepackage{stfloats}
\usepackage{xcolor}
\usepackage{pgfplots}
\usepgfplotslibrary{groupplots}       % enable group‐plot functionality
\usepackage{amsmath}
\usepackage{amssymb}
\usepackage{siunitx}                  % for consistent number formatting, units, etc.
\usepackage{todonotes}
\usepackage{url}
\usepackage{float}
\usepackage{verbatim}
\usepackage{graphicx}
\def\BibTeX{{\rm B\kern-.05em{\sc i\kern-.025em b}\kern-.08em
    T\kern-.1667em\lower.7ex\hbox{E}\kern-.125emX}}
\usepackage{balance}
\usepackage{float}
\usepackage{hyperref}
\usepackage[normalem]{ulem}
\hypersetup{
    colorlinks=true,
    linkcolor=blue,
    filecolor=magenta,      
    urlcolor=blue,
    }

\usepackage[
backend=biber,
sorting=none,
doi=false,
isbn=false,
url=true,
maxcitenames=2,
mincitenames=1,
style=ieee,
]{biblatex}

\usepackage{booktabs}   % enables \toprule, \midrule, \bottomrule
\newcommand{\good}[1]{\textcolor{green!60!black}{#1}}
\newcommand{\bad}[1]{\textcolor{red!70!black}{#1}}

\begin{document}
\title{Context-aware Simopt-Power: Using structural data with simulation metadata to optimise FPGA designs}
\author{\IEEEauthorblockN{Eashan Wadhwa, Georgios Floros \& Shanker Shreejith}
\IEEEauthorblockA{Department of Electronic and Electrical Engineering,
Trinity College Dublin\\
Dublin, Ireland\\
\{wadhwae, florosg, shreejith.shanker\}@tcd.ie}}

\maketitle
\begin{abstract}
Pre-implementation behavioural simulation routinely validates functional correctness, yet it also produces rich switching-activity traces that are typically discarded by FPGA computer-aided design (CAD) flows. Prior simulation-guided and power-aware FPGA optimisations demonstrate the promise of exploiting this metadata, but many rely on fixed thresholds, narrow decision heuristics, or limited design awareness, often incurring substantial area overhead. This paper presents Context-aware Simopt-Power, a simulator-guided optimisation framework that combines activity metadata with lightweight structural features (sequential proximity, logic-depth proxies, and fan-out estimates) to more precisely target high-impact regions of the netlist. We additionally remove empirically tuned constants, replacing them with architecture-aware parameters such as LUT size and mapping constraints, and evaluate trade-offs using power, delay, and a more useful metrics, area-delay product (AD) and power-delay product (PD). Implemented in an open-source Yosys/ABC flow and evaluated on the complex Koios deep-learning accelerator benchmarks, Context-aware Simopt-Power achieves an average \textbf{6.8\%} dynamic-power reduction while limiting LUT overhead to \textbf{11.2\%}, thus enabling a holistic design optimisation. 
%Relative to prior simulation-guided baselines with the original Simopt-Power serving as our direct implementation baseline, these results provide a more favourable balance between power reduction and resource utilisation, supporting more systematic and portable switching-activity optimisation.
\end{abstract}
%\todo[inline]{I think we are being very narrow in comparison by specifying Simopt directly; a broader comparison against power-aware optimisers would be better here. \textcolor{green}{done}}

\section{Introduction}
FPGAs have achieved widespread adoption across different domains from data centres to edge devices, and embedded systems, primarily as they enable power efficient design implementations compared to off-the-shelf platforms. Among the contributors to FPGA power consumption, dynamic power, dominated by signal switching activity across the highly configurable logic and interconnect fabric, remains the most significant and difficult to control. Frequent toggling of wide datapaths and global routing resources amplifies capacitive charging losses, making switching activity a critical lever for reducing overall energy consumption without altering device voltage or frequency.
Traditional FPGA CAD flows primarily focus on functional correctness and timing closure, often overlooking the rich switching activity data available from pre-implementation simulation. While behavioural simulation is routinely employed to verify functional correctness, it also yields detailed toggle-rate and timing metadata that conventional FPGA CAD flows leave unused. This metadata, if harnessed effectively, can inform targeted optimisations to reduce dynamic power consumption as proven in prior works \cite{11231274}.
However, integrating simulation-derived insights into synthesis and mapping remains challenging because modern FPGA fabrics are heterogeneous and routing-dominated, while the CAD flows are not natively optimised around behavioural activity metadata.
Prior work has explored switching-activity reduction through switching-graph constraints \cite{kubica2024switching}, clock-gating approaches \cite{hameed2022power}\cite{xue2025regate}, and other power-oriented gating heuristics \cite{vaithianathan2024low}. These methods show the value of power-aware optimisation, but they do not tightly couple simulation metadata with structural cues for switching-aware FPGA restructuring. This has proven to be very beneficial - \cite{3670474}  leverages structural design representations to guide design-space exploration in high-level synthesis for latency designs. Similarly, \cite{electronics13091744} integrates synthesis-aware modelling while jointly considering multiple performance indicators during neural-network accelerator prototyping. Collectively, these works highlight the benefits of embedding structural awareness into optimisation workflows to achieve more balanced and effective design trade-offs.
In \cite{11231274}, the authors exploit a simulation-driven switching-reduction technique and reported measurable power savings on sequential benchmarks. However, its implementation depends on fixed thresholds and ad hoc heuristics, leading to coarse optimisation and substantial area overhead. The absence of explicit structural awareness in the decision process also limits how effectively it can balance power reduction against downstream resource cost.
In this work, we present \textbf{Context-aware Simopt-Power}, an enhanced framework that couples simulation-derived activity metadata with structural design information to enable more principled switching-activity optimisation. Our main contributions are as follows:
%\todo[inline] {Since this is a 4 page paper, better to combine into \& previous works into one section (Intro \& Background) \textcolor{green}{done}}

\begin{itemize}
\item \textbf{Context-aware Shannon splitting heuristic:} We extend the existing truth-table-based decomposition strategy with lightweight structural features, such as sequential proximity, logic depth proxies, and fan-out estimates to guide optimisation decisions. This enables targeted transformations within deep inter-register logic, guard-dominated regions, and wide, high-activity cones where switching suppression yields the greatest benefit.

\item \textbf{Elimination of heuristic constants:} We remove empirically tuned thresholds and replace them with architecture-aware parameters (such as LUT sizing and mapping constraints), improving portability across FPGA families though systematic optimisation.

\item \textbf{Area-Delay and Power-Delay metrics:} We adopt a more comprehensive evaluation metric that captures the joint impact of power, area and performance, providing a holistic metric for analysing trade-offs in overall design efficiency.
\end{itemize}

\section{Implementation}
%\todo[inline]{We need to check if double blinding is required for the conference, if this is the case, we will need to remove any pointers towards simopt, as it will directly identify the authors and disqualify the submission.}
We build on the Shannon-decomposition-based truth-table splitting routine, described in~\cite{11231274}, as the fundamental transformation, but refine its decision policy using structural features and eliminate fixed, hand-tuned thresholds. Simopt framework \cite{simopt} first collects per-net toggle statistics during behavioural simulation, which serve as switching-activity estimates to highlight high-activity cones. %, previously shown in \cite{11231274} to be strong candidates for power-oriented restructuring. 
Based on these estimates, we extract candidate cuts and apply activity guided transformations within specific logic cones from ABC's (Yosys) internal AIG representation and cut-enumeration data structures, prior to final LUT binding. 
%
%The design is synthesised and technology-mapped to a LUT-based architecture via Yosys with the \textit{ABC} back end. From ABC's internal AIG representation and cut-enumeration data structures, we extract candidate cuts and apply activity-guided transformations within the corresponding logic cones prior to final LUT binding. 
Specifically, the optimisations we introduce comprise: (i) context-aware Shannon cofactoring of high-toggle variables within selected cuts; (ii) selective logic duplication to localise switching and reduce glitch propagation; (iii) restructuring of wide, guard-dominated cones to reduce unnecessary reconvergence; and (iv) activity-weighted cut re-evaluation to favour implementations that lower estimated dynamic power while respecting LUT size constraints. The modified network is then returned to ABC's standard mapping flow to produce the final netlist. Each of these optimisation components are described in the following subsections. Alg.~\ref{alg:ttsplit_new} provides the pseudocode for the ABC-integrated context-aware Shannon splitting routine which we call \textsc{TTContextualDecompose}. This algorithm evaluates a candidate cut and optionally applies Shannon-style cofactoring when the context-aware cost justifies the LUT overhead. . 

%removed explicit callouts that repeat the info -- - LUT aware gating thresholds, expected-value model for sequential guard evaluation, overhead penalty scaled by truth-table size, and mux sensitivity penalty from cofactor difference -- 

\subsection{Structural eligibility filter for deep sequential cones}\label{sec_structural_filter}
\begin{figure}[h]
  \centering
  \includegraphics[width=0.42\textwidth]{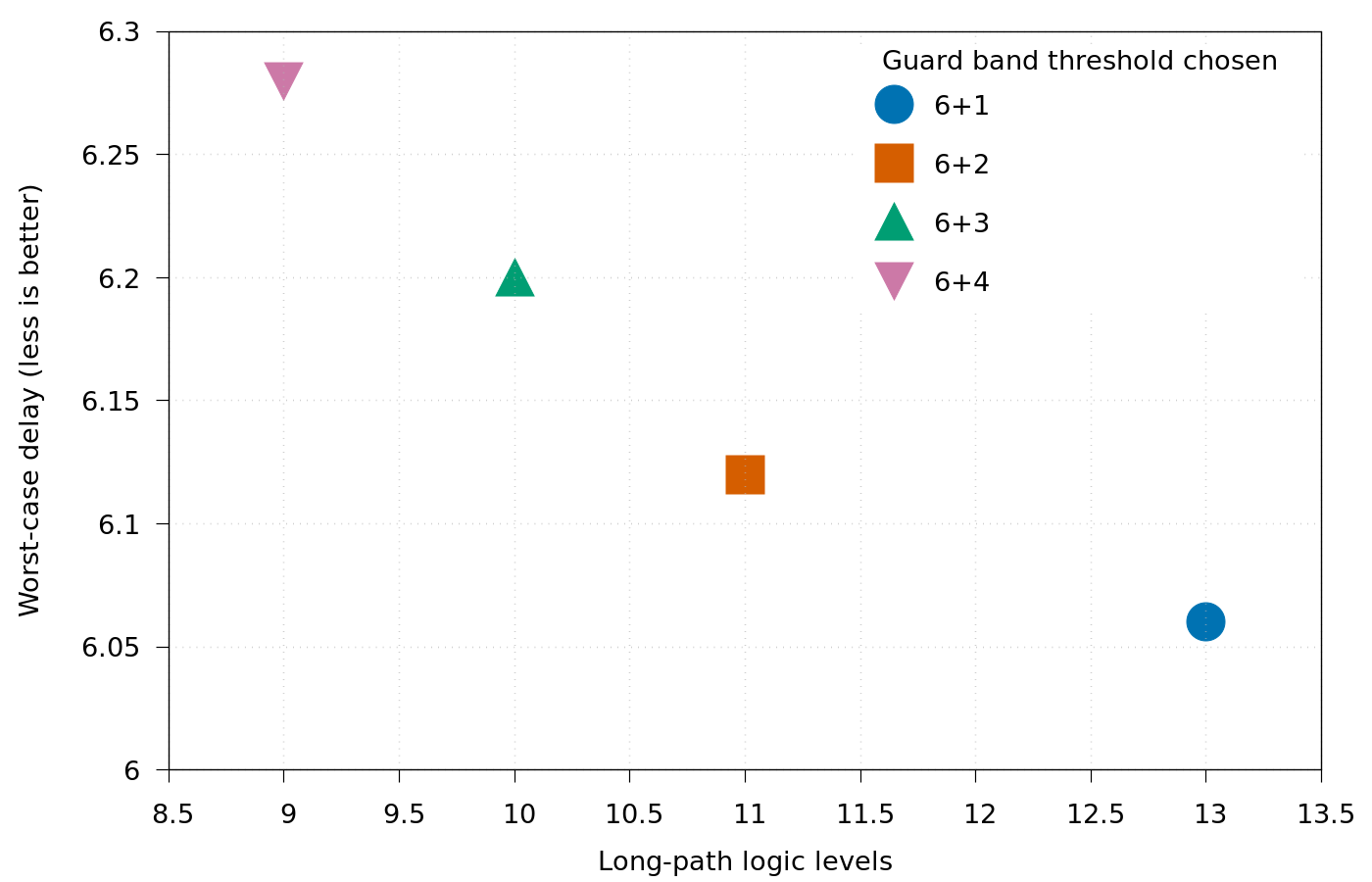}
  \caption{A sweep scatter of various guard-band thresholds of worst-case delays vs logic levels on a critical path. $+2$ is the sweet spot for this architecture chosen with $K=6$}
  \label{fig:1} \vspace{-5mm}
\end{figure}

%\todo[inline]{We have to define what is \textsc{TTContextualDecompose} somewhere. \textcolor{green}{done}}
We first constrain \textsc{TTContextualDecompose} to prioritise cuts whose right-cut function ($\mathsf{T}$) resides within deep, serialised combinational regions. 
In the ABC internal representation, a logic cone is classified as "large logic between flops" when it is part of a sequential path and its estimated cut depth ($d$) exceeds a LUT-aware minimum threshold. The depth metric is exposed through an augmented mapping-context variable within ABC.
This restriction is motivated by dynamic-power considerations: deep combinational regions are evaluated every clock cycle and typically exhibit numerous internal nodes and reconvergent paths. Thus reducing switching at the cone output is more likely to propagate meaningful activity suppression throughout the surrounding datapath. Moreover, these regions better amortise the area overhead introduced by Shannon cofactoring. To ensure architectural portability, we avoid hard-coded constants and instead define ($K := \max(\texttt{lutSize}, 6)$) and ($d_{\min} := K + 2$). $K$ denotes the effective LUT input size of the target FPGA architecture, derived from the mapper's \texttt{lutSize} parameter with six set as the lower bound to ensure architectural portability. A $+2$ guard-band threshold was chosen, to ensure the sequential cones are sufficiently deep so that the expected switching reduction can amortise the split overhead when compared to other threshold values shown in Fig.\ref{fig:1}. The split search is therefore enabled only when the condition (is sequential and the condition $(d \ge d_{\min})$) holds, otherwise, \textsc{TTContextualDecompose} terminates early without modifying ($\mathsf{T}$). 

\subsection{Expected-activity model for sequential guard evaluation}\label{sec_expected_activity}
  
In~\cite{11231274}, the authors computed Shannon cofactors by enumerating the $2^n$ minterms of a truth table, where $n$ is the truth-table size of the cut function. While exact, this brute-force view treats all cuts uniformly and does not exploit any per-cut context (e.g., whether the cone is sequential, or whether a split variable is likely to behave as a guard). To mitigate this, we introduce an \textit{expected-ones-count model} in \textsc{TTContextualDecompose} that estimates the activity of a Shannon split using a probabilistic prior, rather than assuming the quieter cofactor is always selected.

Let $\mathsf{T}$ denote the packed truth table of a cut on $n$ variables, and let $N_1(\mathsf{T})$ be the number of minterms set to 1 in $\mathsf{T}$ (popcount of the truth-table bits). For a candidate split variable at position $pos$, we form the Shannon cofactors $\mathsf{T}_0$ and $\mathsf{T}_1$ by fixing the variable to 0 and 1, respectively. Define $N_0 := N_1(\mathsf{T}_0)$ and $N_1 := N_1(\mathsf{T}_1)$, and let $N_{\min} := \min(N_0,N_1)$ and $N_{\max} := \max(N_0,N_1)$. In prior optimisations, the split was implicitly modelled as if the quieter (i.e. the lower simopt-counter count) branch is always selected (i.e., the post-split ones-count equals $N_{\min}$). We instead model post-split activity by an expected value: %\todo{what is a quiet branch? \textcolor{green}{done}}
\begin{equation}
  \mathbb{E}\!\left[N_{\mathrm{after}}\right]
  :=
  \pi\,N_{\min} + (1-\pi)\,N_{\max},
  \label{eq:expected_after}
\end{equation}
where $\pi \in [0.5,0.95]$ is a prior probability that the \emph{quiet} branch (the cofactor achieving $N_{\min}$) is taken at runtime. For purely combinational cones we set $\pi = 0.5$ (no bias), while for sequential cones we derive $\pi$ from the Simopt score $s\in[0,s_{max}]$, where $s_{max}$ is the second highest Simopt score observed after the clock nets.
\begin{equation}
  \pi(s) :=
  \mathrm{clip}\!\left(0.95 - 0.45\cdot\frac{s}{s_{max}},\ [0.5,0.95]\right),
  \label{eq:pi_score}
\end{equation}
so that higher scores revert toward an unbiased (random) prior and lower scores permit a stronger guard-like bias toward the quiet branch. Finally, we define the expected improvement in ones-count as
\begin{equation}
  \Delta N := N_1(\mathsf{T}) - \mathbb{E}\!\left[N_{\mathrm{after}}\right],
  \label{eq:deltaN}
  \end{equation}
which is used as a switching-reduction proxy when ranking split candidates. This prevents the splitting of logic nets that are already low-activity, the split is only selected when the expected reduction in output activity ($\Delta N$) is positive and large enough to amortise the estimated mux/duplication overhead. This component is primarily intended to model guard-dominated behaviour in deep sequential or glitch-prone cones.
  % We primarily achieved this through logging (s, N_before,
  % N_min, N_max) from real ttSplit candidate evaluations or selected splits, then plot E[N_after] computed from those logged values

\subsection{LUT-aware overhead modelling}\label{sec_lut_overhead}
\begin{figure}[h]
  \centering
  \includegraphics[width=0.42\textwidth]{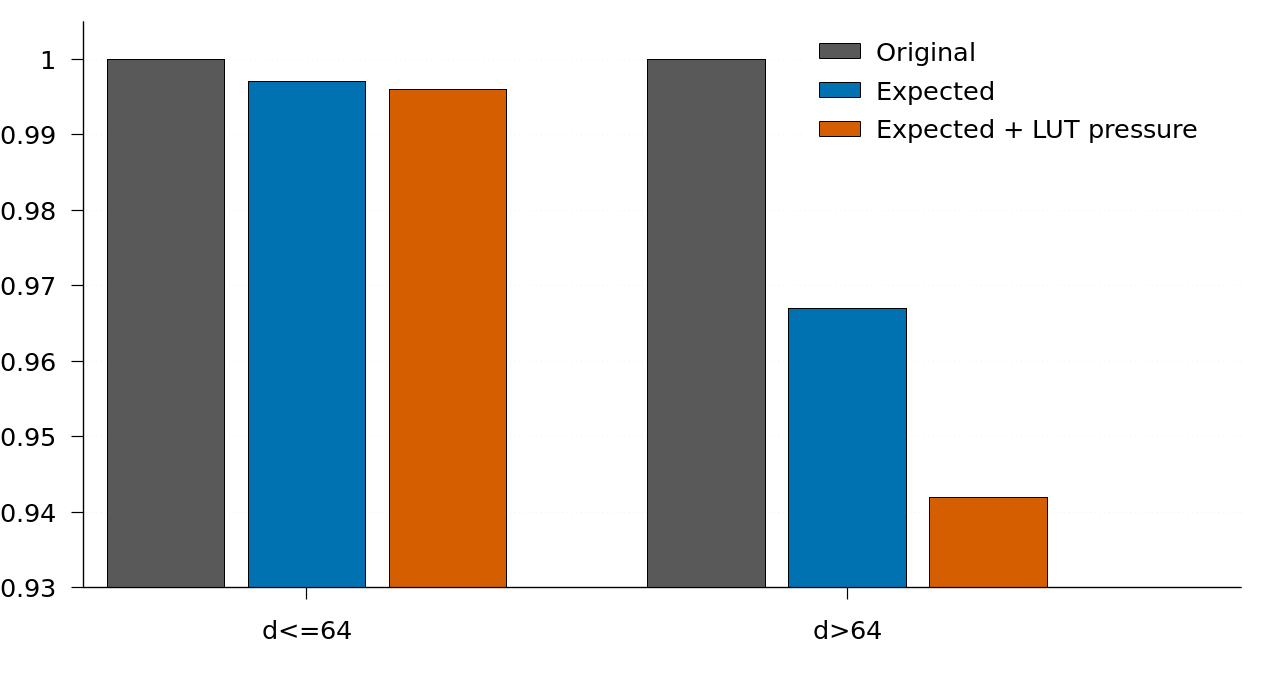}
  \caption{Impact to area with Expected-value model optimisation (shown by the middle bar) and the Expected-value model with $P_{\mathrm{lut}}$ optimisations (the right-most bar) for lstm.v. The x-axis has two regimes: shallow cones ($d<=64$) and deep cones ($d>64$), where $d$ corresponds to the estimated sequential cone depth. Y-axis represents the normalised area impact.} \vspace{-5mm}
  \label{fig:2}
\end{figure}

While Shannon decomposition can reduce switching activity, it also introduces structural overhead in the form of two cofactors and a selection function (conceptually, a mux). To avoid favouring splits that are unlikely to map efficiently, we incorporate a LUT-capacity term that reflects packing pressure. Let $n_{\text{new}} := |\mathcal{C}_{\text{new}}|$ denote the support size of the candidate cut and let $K$ be the target LUT input capacity. When $n_{\text{new}}$ approaches $K$, any additional logic created by the split is harder to absorb within a single LUT and is more likely to increase LUT count; conversely, when $n_{\text{new}} \ll K$, the mapping has slack and the overhead is typically easier to pack. We therefore apply a LUT-pressure factor $P_{\mathrm{lut}}$ that increases the overhead penalty as $n_{\text{new}}$ nears (or exceeds) $K$, biasing the search towards decompositions that either (i) split near-capacity functions into smaller-support cofactors that map more cleanly, or (ii) avoid splits whose
overhead would likely spill into additional LUTs. The illustration of the impact of these two parameters is shown in Fig.~\ref{fig:2}, where large area savings are possible for large cone depths. 

\subsection{Cofactor divergence penalty}\label{sec_cofactor_divergence}
If the two  cofactors $\mathsf{T}_0$ and $\mathsf{T}_1$ are very large, then the decomposed output depends strongly on the split variable: toggling the split variable can cause many output minterms to change, which is undesirable when the goal is to reduce activity. We capture this effect by computing the XOR of the cofactors and counting how many minterms differ between them. Subseqeently, we define a sensitivity penalty \[ P_{\mathrm{sens}} := \cdot N_1(\mathsf{T}_0 \oplus \mathsf{T}_1), \] where $N_1(\cdot)$ is the number of ones (i.e., differing minterms) in the XOR truth table. This term discourages split variables that behave like a highly sensitive selector rather than a guard.

\subsection{Final split score and decision rule}
The components described above are combined into a single scalar score that balances expected switching reduction against structural penalties.  Before evaluating any split candidates, we first apply the structural eligibility filter described in Section~\ref{sec_structural_filter}. The contextual decomposition is attempted only if the cut lies in a sufficiently deep sequential cone, i.e. $(d \ge d_{\min})$where $d_{\min} = K + 2$ and $K$ is the effective LUT size. If this condition does not hold, the truth table $\mathsf{T}$ is left unchanged and no further evaluation is performed.
For a candidate split variable, we compute the expected reduction in ones-count
$\Delta N$ (Eq.~\ref{eq:deltaN}) from Section~\ref{sec_expected_activity}, the LUT-pressure penalty
$P_{\mathrm{lut}}$ (from Section~\ref{sec_lut_overhead}), and the mux-sensitivity penalty
$P_{\mathrm{sens}}$ (from Section~\ref{sec_cofactor_divergence}). The overall score is defined as
\begin{equation}
\mathrm{score}(\textit{splitVar})
:=
\Delta N
\;-\;
P_{\mathrm{lut}}
\;-\;
P_{\mathrm{sens}}.
\label{eq:finalscore}
\end{equation}
We summarise this algorithm change in Algorithm~\ref{alg:ttsplit_new}. This new function is stitched into the existing Simopt Yosys-ABC flow at the cut-processing stage of technology mapping, where each candidate cut is evaluated. Once all cuts have been processed, ABC emits the mapped BLIF netlist, which can be imported into subsequent place and route stages. For our evaluation, we use Vivado to target FPGA implementations. 
%is then imported into Vivado as a black-box implementation of the optimised design and passed through the post-synthesis power and timing analysis flow. 
We discuss this further in Sec.~\ref{section:result}

\begin{algorithm}[t]
\scriptsize
\caption{\textsc{TTContextualDecompose}}
\label{alg:ttsplit_new}

% \KwIn{Cut $C$ with truth table $\mathsf{T}$}
% \KwOut{Possibly modified truth table}

\If{not $(\texttt{isSequential} \land d \ge d_{\min})$}{
    \Return $\mathsf{T}$ \tcp*{Structural eligibility not satisfied}
}

\ForEach{candidate variable $x \in C$}{
    compute $\Delta N$ \tcp*{Expected switching reduction from guard model}
    compute $P_{\mathrm{ovh}}$ \tcp*{Estimated overhead from LUT-aware model}
    compute $P_{\mathrm{sens}}$ \tcp*{Penalty from cofactor divergence}  
    $score \gets \Delta N - P_{\mathrm{ovh}} - P_{\mathrm{sens}}$\;
    
    \If{$score > bestScore$}{
        $bestScore \gets score$\;
        $bestVar \gets x$\;
    }
}

\If{$bestScore > 0$}{
    \Return \textsc{TTDecompose}$(C, bestVar)$ \tcp*{Without splitting logic from \cite{11231274}}
}
\Else{
    \Return $\mathsf{T}$\;
}

\end{algorithm}

% \todo{should we not specify how this is finally mapped back into the synthesis flow using one or two lines? \textcolor{green}{done}}
\section{Results}\label{section:result}
To evaluate the proposed enhancements in the Context-Aware Simopt-Power (C.A.) framework, we compared it against the baseline Simopt-Power (S.P.) implementation (from~\cite{11231274}) using the Koios benchmark \cite{Koios}. This benchmark comprises deep-learning circuits of varying complexity, with inputs that are compatible with Verilator, which we use to generate the toggle data from (Simopt-counters). 
We measured dynamic power reduction, LUT overhead, and area-delay-product to assess the trade-offs arising from context-aware Simpopt-Power using reports from the post-place-and-route flow. 
%\todo[inline]{Replace the bit below with the Yosys + Vivado flow -- Post logic-optimisation through Simopt, we use Yosys-ABC to generate the netlist, which is fed to Vivado for placement and routing targeting X FPGA. The effective Area considered is the LUT utilisation, while Delay is captured from the post-place and route timing reports. Vectored power estimation is used to generate power consumption estimates using Vivado's estimator and activity generated from simulation \textcolor{green}{done}}.
For place and route, the optimised netlist post C.A. and standard S.P. are exported as (\textit{*.blif}) into Vivado, with \textit{xc7a200tsbv484\-1} as the target device to generate the implemented design. 
%To get these metrics, we import the mapped netlist generated (\textit{*.blif}) from C.A. Yosys-ABC framework into Vivado for getting the powers and timing numbers. We chose \textit{xc7a200tsbv484\-1} as the target FPGA when generating the Vivado reports. 
Area is captured as LUT utilisation from the implementation reports, whereas delay represents the critical path delay from the post-place-and-route timing reports. 
Power is estimated using Vivado's vectored power analysis with switching activity (\textit{*.saif}) as input obtained from Verilator. 
The area, power and delay gains/overheads are compared against the baseline implementation of the circuit through standard Vivado implementation flow.  
%For the baseline designs, dynamic power is taken directly from the Vivado power reports, while the S.P. and C.A. columns (indicated as CA in table) apply the corresponding percentage changes shown in the table relative to these baseline values, with the LUT counts kept fixed.   
% \todo[inline]{You are using CA in the table, so stick with CA for context aware in the text. \textcolor{green}{done, is C.A. and S.P.}}
As shown in table~\ref{tab:koios_sp_apd}, our context-aware flow achieves an average dynamic power reduction of \(\approx\)\textbf{6.8\,\%} across the benchmark, which is a slight reduction over the original S.P.'s \(\approx\)\textbf{9.4\,\%} power gain, compared to the baseline implementation. However, we observe a significant improvement in LUT overhead, with C.A. flow incurring only \(\approx\)\textbf{11.21\,\%} average increase in LUT area compared to the original S.P.'s \(\approx\)\textbf{18.63\,\%} increase, both measured over the baseline implementation of the circuit (w/o column in table). This improvement is better reflected in the percentage changes in the Power-Delay (PD) product and the Area-Delay (AD) product with respect to the baseline implementation (captured by $\Delta$PD and $\Delta$AD columns respectively). We can see that C.A. achieves nearly the same energy efficiency (PD) as the S.P. implementation at significantly better area efficiency (AD) across all designs, demonstrating its ability for performance favourable trade-off between power reduction and resource utilisation. This can be attributed to the targeted and context-aware nature of the optimisations in C.A., which allows for more efficient use of LUT resources while still achieving meaningful power savings. Overall, the results indicate that our context-aware enhancements provide a more balanced optimisation strategy, delivering respectable power reductions with significantly lower area overhead, thus improving the overall efficiency of the FPGA designs.

% \textcolor{red}{This improvement in area efficiency is reflected in the area-power-delay-product (ADP) metric, where CA achieves an average ADP improvement of \(\approx\)\textbf{2.85\,\%} compared to the original Simopt-Power's \(\approx\)\textbf{6.78\,\%} increase, demonstrating a more favourable trade-off between power reduction and resource utilisation.} This is due to the more targeted and context-aware nature of the optimisations, which allows for more efficient use of LUT resources while still achieving meaningful power savings. Overall, the results indicate that our context-aware enhancements provide a more balanced optimisation strategy, delivering respectable power reductions with significantly lower area overhead, thus improving the overall efficiency of the FPGA designs.
% \todo{inline}{Come back and fix the text here, once the table is clarified}

% \todo[inline]{I think this table needs to be rethought, particularly for the delta numbers -- what are we indicating here? Is higher better? The green red tagging is a bit confusing as I see bigger greens with SP over CA for final APD -- is that expected? }
\begin{table*}[t]
\centering
\caption{Power-area-delay comparison for Simopt-Power (S.P.) and Context-Aware Simopt-Power (C.A.). AD = Area$\cdot$Delay, PD = Power$\cdot$Delay.}
\label{tab:koios_sp_apd}
\resizebox{\textwidth}{!}{%
\begin{tabular}{l
rrrrr
rrrrr
rrr
rrrrr
rrrrr}
\toprule
Benchmark
& \multicolumn{3}{c}{Power} & \multicolumn{2}{c}{$\Delta P$}
& \multicolumn{3}{c}{Area} & \multicolumn{2}{c}{$\Delta A$}
& \multicolumn{2}{c}{Delay} & $\Delta D$
& \multicolumn{3}{c}{AD} & \multicolumn{2}{c}{$\Delta$AD (\%)}
& \multicolumn{3}{c}{PD} & \multicolumn{2}{c}{$\Delta$PD (\%)} \\
\cmidrule(lr){2-4}\cmidrule(lr){5-6}
\cmidrule(lr){7-9}\cmidrule(lr){10-11}
\cmidrule(lr){12-13}\cmidrule(lr){14-14}
\cmidrule(lr){15-17}\cmidrule(lr){18-19}
\cmidrule(lr){20-22}\cmidrule(lr){23-24}

& w/o & w/S.P. & w/C.A.
& S.P. & C.A.
& w/o & w/S.P. & w/C.A.
& S.P. & C.A.
& w/o & w/S.P.,C.A.
& S.P.,C.A.
& w/o & w/S.P. & w/C.A.
& S.P. & C.A.
& w/o & w/S.P. & w/C.A.
& S.P. & C.A. \\
\midrule

dla\_like
& 1.86 & 1.69 & 1.73 & \good{8.8} & \good{6.7}
& 479619 & 571931 & 533264 & \bad{19.2} & \bad{11.2}
& 35.10 & 34.71 & \good{1.1}
& 16834627 & 19851725 & 18509593 & \bad{17.9} & \bad{9.9}
& 65.3 & 58.7 & 60.0 & \good{10.1} & \good{8.0} \\

clstm\_like
& 0.93 & 0.86 & 0.87 & \good{8.3} & \good{6.5}
& 201945 & 235509 & 222978 & \bad{16.6} & \bad{10.4}
& 31.14 & 30.76 & \good{1.2}
& 6288567 & 7244257 & 6858803 & \bad{15.2} & \bad{9.1}
& 29.0 & 26.5 & 26.8 & \good{8.7} & \good{7.6} \\

deepfreeze
& 0.45 & 0.41 & 0.42 & \good{8.9} & \good{6.9}
& 75729 & 83302 & 81999 & \bad{10.0} & \bad{8.3}
& 70 & 71 & \bad{1.4}
& 5301030 & 5914442 & 5821929 & \bad{11.6} & \bad{9.8}
& 31.5 & 29.1 & 29.8 & \good{7.6} & \good{5.3} \\

tdarknet\_like
& 0.95 & 0.87 & 0.89 & \good{8.4} & \good{6.5}
& 159873 & 199841 & 181392 & \bad{25.0} & \bad{13.5}
& 78 & 79 & \bad{1.3}
& 12470094 & 15787439 & 14329968 & \bad{26.6} & \bad{14.9}
& 74.1 & 68.7 & 70.3 & \good{7.2} & \good{5.1} \\

bwave\_like
& 0.08 & 0.068 & 0.071 & \good{9.7} & \good{7.2}
& 521691 & 652114 & 591728 & \bad{25.0} & \bad{13.4}
& 14.11 & 13.96 & \good{1.1}
& 7361060 & 9103511 & 8260523 & \bad{23.7} & \bad{12.2}
& 1.1 & 0.9 & 1.0 & \good{15.9} & \good{12.2} \\

lstm
& 2.02 & 1.84 & 1.88 & \good{8.9} & \good{6.8}
& 274477 & 307414 & 299029 & \bad{12.0} & \bad{8.9}
& 27.04 & 26.72 & \good{1.2}
& 7421858 & 8214102 & 7990055 & \bad{10.7} & \bad{7.7}
& 54.6 & 49.2 & 50.2 & \good{10.0} & \good{8.0} \\

bnn
& 0.839 & 0.774 & 0.787 & \good{7.7} & \good{6.2}
& 43754 & 51630 & 48510 & \bad{18.0} & \bad{10.9}
& 31.09 & 30.59 & \good{1.6}
& 1360312 & 1579362 & 1483921 & \bad{16.1} & \bad{9.1}
& 26.1 & 23.7 & 24.1 & \good{9.2} & \good{7.7} \\

lenet
& 0.012 & 0.011 & 0.011 & \good{8.6} & \good{6.4}
& 23560 & 30392 & 27045 & \bad{29.0} & \bad{14.8}
& 20.90 & 20.52 & \good{1.8}
& 492404 & 623644 & 554963 & \bad{26.7} & \bad{12.7}
& 0.3 & 0.2 & 0.2 & \good{10.0} & \good{10.0} \\

dnnweaver
& 1.50 & 1.35 & 1.39 & \good{10.0} & \good{7.3}
& 252431 & 323429 & 288705 & \bad{28.1} & \bad{14.4}
& 84 & 85 & \bad{1.2}
& 21204204 & 27491465 & 24539925 & \bad{29.7} & \bad{15.7}
& 126.0 & 114.8 & 118.1 & \good{8.9} & \good{6.2} \\

tpu\_like
& 0.094 & 0.086 & 0.088 & \good{8.5} & \good{6.6}
& 597420 & 704956 & 662778 & \bad{18.0} & \bad{10.9}
& 73.87 & 73.06 & \good{1.1}
& 44131415 & 51504085 & 48422561 & \bad{16.7} & \bad{9.7}
& 6.9 & 6.3 & 6.4 & \good{9.5} & \good{7.4} \\

gemm\_layer
& 0.062 & 0.056 & 0.058 & \good{8.9} & \good{6.8}
& 104338 & 125639 & 118236 & \bad{20.4} & \bad{13.3}
& 73 & 74 & \bad{1.4}
& 7616674 & 9297286 & 8749464 & \bad{22.1} & \bad{14.9}
& 4.5 & 4.1 & 4.3 & \good{8.4} & \good{5.2} \\

attention\_layer
& 2.20 & 1.98 & 2.04 & \good{10.0} & \good{7.3}
& 353403 & 392277 & 383531 & \bad{11.0} & \bad{8.5}
& 16.30 & 16.12 & \good{1.1}
& 5760469 & 6323505 & 6182520 & \bad{9.8} & \bad{7.3}
& 35.9 & 31.9 & 32.9 & \good{11.0} & \good{8.3} \\

conv\_layer
& 0.042 & 0.039 & 0.040 & \good{7.3} & \good{6.0}
& 137995 & 155934 & 150532 & \bad{13.0} & \bad{9.1}
& 10.21 & 10.08 & \good{1.3}
& 1408929 & 1571815 & 1517363 & \bad{11.6} & \bad{7.7}
& 0.4 & 0.4 & 0.4 & \good{8.3} & \good{6.0} \\

robot\_rl
& 1.88 & 1.72 & 1.76 & \good{8.8} & \good{6.5}
& 28608 & 31469 & 30897 & \bad{10.0} & \bad{8.0}
& 24.80 & 24.38 & \good{1.7}
& 709478 & 767214 & 753269 & \bad{8.1} & \bad{6.2}
& 46.6 & 41.9 & 42.9 & \good{10.1} & \good{8.0} \\

reduction\_layer
& 0.13 & 0.11 & 0.12 & \good{10.9} & \good{8.2}
& 18511 & 20952 & 20180 & \bad{13.2} & \bad{9.0}
& 16.20 & 15.89 & \good{1.9}
& 299878 & 332927 & 320660 & \bad{11.0} & \bad{6.9}
& 2.1 & 1.7 & 1.9 & \good{17.0} & \good{9.5} \\

spmv
& 0.136 & 0.123 & 0.126 & \good{10.0} & \good{7.5}
& 13463 & 16425 & 15120 & \bad{22.0} & \bad{12.3}
& 18.17 & 18.17 & \good{0.0}
& 244623 & 298442 & 274730 & \bad{22.0} & \bad{12.3}
& 2.5 & 2.2 & 2.3 & \good{9.6} & \good{7.4} \\

eltwise\_layer
& 0.042 & 0.038 & 0.039 & \good{9.5} & \good{7.4}
& 15987 & 19344 & 17915 & \bad{21.0} & \bad{12.1}
& 9.85 & 9.66 & \good{1.9}
& 157472 & 186863 & 173059 & \bad{18.7} & \bad{9.9}
& 0.4 & 0.4 & 0.4 & \good{11.3} & \good{8.9} \\

softmax
& 0.49 & 0.43 & 0.45 & \good{10.8} & \good{7.7}
& 10938 & 13126 & 12223 & \bad{20.0} & \bad{11.7}
& 39.05 & 39.05 & \good{0.0}
& 427129 & 512570 & 477308 & \bad{20.0} & \bad{11.7}
& 19.1 & 16.8 & 17.6 & \good{12.2} & \good{8.2} \\

conv\_layer\_hls
& 0.98 & 0.90 & 0.91 & \good{8.3} & \good{6.5}
& 121167 & 153882 & 138112 & \bad{27.0} & \bad{14.0}
& 20.66 & 20.39 & \good{1.3}
& 2503310 & 3137654 & 2816104 & \bad{25.3} & \bad{12.5}
& 20.2 & 18.4 & 18.6 & \good{9.4} & \good{8.4} \\

proxy
& 0.055 & 0.050 & 0.051 & \good{9.1} & \good{7.3}
& 9255 & 10551 & 10131 & \bad{14.0} & \bad{9.5}
& 45 & 45 & \good{0.0}
& 416475 & 474795 & 455895 & \bad{14.0} & \bad{9.5}
& 2.5 & 2.2 & 2.3 & \good{9.1} & \good{7.3} \\

\bottomrule
\end{tabular}}
\end{table*}
% \todo[inline]{Are we missing a few references?}
\section{Conclusion}
In this work, we presented \textbf{Context-aware Simopt-Power}, an enhanced framework for simulation-driven switching-activity optimisation in FPGA design flows. By integrating structural design information with simulation-derived activity metadata, we developed a more principled and targeted approach to reducing dynamic power consumption. Our context-aware Shannon splitting heuristic, elimination of heuristic constants, and comprehensive area-delay-product evaluation demonstrated improved trade-offs between power reduction and resource utilisation compared to the original Simopt-Power implementation. Future work will explore further refinements to the decision policy, such as incorporating additional structural features and extending the design choices to be more data and workload-aware.

\printbibliography

\end{document}